\documentclass[11pt,oneside,a4paper]{article}

\usepackage{lineno}
\usepackage{hyperref}
\usepackage{graphicx}
\usepackage{latexsym}
\usepackage{amsmath,amscd,amssymb,amsfonts,mathrsfs,amsbsy}
\usepackage{amsthm}
\theoremstyle{definition}

\usepackage{color,xcolor}
\usepackage{enumitem}
\newcommand{\from}{\leftarrow}
\newcommand{\both}{\leftrightarrow}

\newcommand{\mG}{\mathcal{G}}
\newcommand{\mV}{\mathcal{V}}
\newcommand{\mE}{\mathcal{E}}
\newcommand{\mR}{\mathcal{R}}
\newcommand{\mA}{\mathcal{A}}
\newcommand{\mL}{\mathcal{L}}

\newcommand{\sgn}{\text{sign}}
\newcommand{\pa}{par}

\begin{document}
\newtheorem{definition}{Definition}
\newtheorem{theorem}{Theorem}
\newtheorem{proposition}{Proposition}
\newtheorem{lemma}{Lemma}
\newtheorem{example}{Example}

\begin{center}
{\bf\LARGE General theory for stochastic admixture graphs and F-statistics}

\medskip
\medskip
\noindent 
Samuele Soraggi  \footnote{Department of Biology,  University of Copenhagen}, Carsten Wiuf  \footnote{Department of Mathematical Sciences, University of Copenhagen}\footnote{Corresponding author. Email \texttt{wiuf@math.ku.dk}}

\medskip
\medskip
\noindent
July 18, 2018
\end{center}




\begin{center}
--------------------------------------------------------------------------------------------------
\end{center}
{\bf Abstract:} We provide a general mathematical  framework based on the theory of graphical models to study  admixture graphs. Admixture graphs are used to describe  the ancestral relationships between past and present populations, allowing for population merges and migration events, by means of gene flow.
We give various mathematical properties of  admixture graphs with particular focus on properties of the so-called $F$-statistics. Also the Wright-Fisher model is studied and a general expression for the loss of heterozygosity is derived.

\medskip
\noindent
{\bf Keywords:} introgression, genetic drift, Wright-Fisher model, heterozygosity, F-statistic, Markov graphical model. {\bf MSC:} 92D15, 92D25.

\begin{center}
--------------------------------------------------------------------------------------------------
\end{center}

\section{Introduction}
Inference on human demographic history from a genetic perspective has been a topic of wide interest in population genetics \cite{Cavalli-Sforza1966,Cavalli-Sforza1967,Reich2009,Green2010,Patterson2012,Wall2013,Skoglund2015}. Methods for the assessment of migration between populations, and the identification of admixture and splitting events have recently been proposed based on  the study of gene flow and introgression between  populations \cite{Reich2009,Patterson2012,Peter2015}. The postulated  demographic relationships between populations are described by a graph, generally referred to as an admixture graph, where each node represents a population, ancient or extant, and  each directed edge represents an ancestral relationship between two populations. Allele frequencies or other genetic quantities characterising the  populations  are associated to the nodes, thereby creating, what we name, a stochastic admixture graph. The difference in allele frequencies between two nodes quantifies the gene flow between the two populations represented by the nodes.
 
The goal of this paper is to provide a stringent mathematical definition and  treatment  of  stochastic admixture graphs and their properties. We use the theory of graphical models \cite{Frydenberg,LauritzenCausal,Whittaker1990} to develop a general mathematical framework to describe stochastic admixture graphs and their associated variables (such as gene frequencies). In the literature stochastic admixture graphs and their properties are often studied with a particular population genetic model  in mind. We abstract  properties that generalise properties of particular models, using the theory of graphical models. 

Many popular computational tools, building on admixture graphs, quantify similarities in genetic composition between populations by  means of moment statistics, called $F$-statistics   \cite{Patterson2012,Reich2009,Peter2015}.  These tools rely on properties of the $F$-statistics that are not always motivated, but nonetheless have proven  essential  to disentangle  complicated genetic ancestries of populations  \cite{Castelo2006,Patterson2012,treeMix,Lipson}. We deduce general properties of  the $F$-statistics, in particular the $F_2$-statistic, and show that the $F$-statistics  can be decomposed in terms of the admixture paths between  populations. Furthermore, we give conditions under which the $F_2$-statistic is additive (in a sense to be made precise later) and forms a metric on the nodes of the graph. In the final section of the paper, we consider the Wright-Fisher diffusion model and give a general formula for the decline of heterozygosity over time in a stochastic admixture graph.

We  envisage that the general theory developed here and the vast amount of exiting literature on model selection and inference on graphical models will be useful to understand inferential properties of stochastic admixture graphs, what can be done and how. All proofs are given in the appendix.

\section{Admixture graphs}

 We consider labeled graphs  with \emph{directed} and  \emph{undirected} edges, and use the notations $i \both j$ (equivalently $j \both i$) and $i \to j$ (equivalently $j \from i$) for an undirected edge between two nodes $i$ and $j$, and a directed edge from $i$ to $j$, respectively. An edge $i\to j$ is said to be ingoing to $j$ and outgoing of $i$.   The  \emph{parents} of  $j$ is the set $\pa(j) = \{ i \, | \, i \to j \,\,\text{is an edge}\}$. The node $i$ is  a  \emph{child} of $j$ if $j$ is a parent of $i$.

\begin{definition}
An \emph{admixture graph} is an edge labeled graph $\mG = (\mV, \mE,\mL)$ without directed cycles. The triplet consists respectively of the set of nodes, edges and labels. The set of nodes $\mV$ is divided into:
\begin{itemize}
\item[(i)] \emph{roots} $\mR$, nodes without ingoing edges. All pairs of roots, and only these, are connected by an undirected edge,
\item[(ii)] \emph{admixed nodes} $\mA$, nodes with ingoing directed edges,
\item[(iii)] \emph{leaves} $\mA_0\subseteq\mA$, admixed nodes without outgoing directed edges.
\end{itemize}
The label $\alpha_{ij}$ of an edge $i\to j$ is  positive and fulfils
\begin{equation}\label{eq:sum_of_alphas}
\sum\limits_{i \in \pa(j)} \alpha_{ij} = 1.
\end{equation}
An edge between two roots $r_1,r_2\in \mR$ has label $\alpha_{r_1r_2}=1$. 
\end{definition}

In the genetic context, an admixed node, say $j$, represents a population that is a mixture of several populations, such that $100\alpha_{ij}$ percent of the size of population  comes from population $i$, $i\in\pa(j)$.  The roots represents  populations that are ancestral to the other populations in the graph, whereas the ancestral relationship between the root populations are assumed unknown and left unspecified.

\begin{figure}[!htbp]
\centering
\includegraphics[width=10cm]{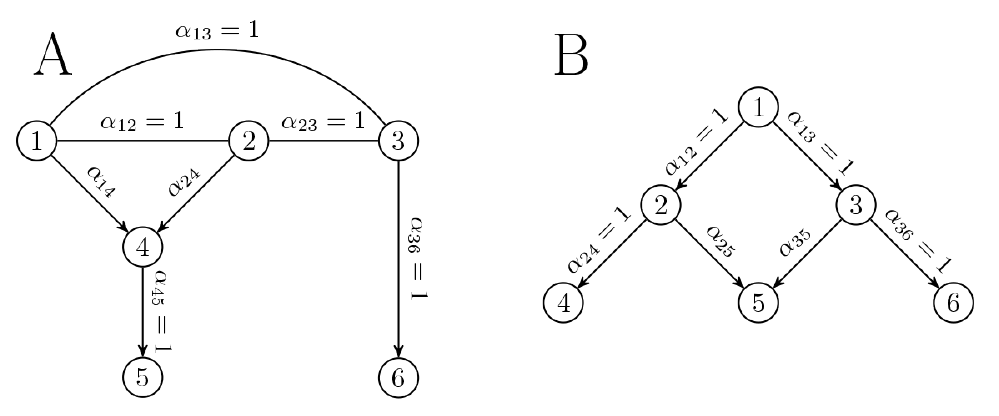}
\caption{{\bf Examples of admixture graphs.} (A) An admixture graph where nodes $1,2,3$ are roots, $4$ an admixed node and $5,6$ a leaves. (B) An admixture graph with three leaves. From $3$ to $5$ there is as an a-path of type (b), namely $(3,5)$ with label $\alpha_{35}$, and one of type (c), namely $(3,1,2,5)$ with label $1\cdot1\cdot \alpha_{36}=\alpha_{36}$.}\label{Fig1}
\end{figure}

By definition, an admixture graph is connected. We assume an admixture graph is not trivial, meaning that it does not consist of only roots and undirected edges. For convenience, we define $\alpha_{ji}=\alpha_{ij}$, $\alpha_{ii}=1$,  and $\alpha_e=\alpha_{ij}$, if $e$ is an edge connecting the nodes $i,j$. See Figure~\ref{Fig1} for examples.

In the following $\mG$  denotes an admixture graph $\mG=(\mV,\mE,\mL)$.

\begin{definition}
\label{defn:path}
Given an admixture graph $\mG$ and two  nodes $i,j\in\mV$, an \emph{admixture path} (or simply an \emph{a-path}) $\gamma$ from $i$ to $j$ is a sequence of  edges with distinct nodes,
\begin{equation*}
i_k\from i_{k-1},\ldots,i_1\from i_0, \qquad j_0\to j_1,\ldots, j_{m-1}\to j_{m},
\end{equation*}
where  $i_k=i,\,j_{m}=j$, $k,m\geq 0$, and $i_0 \neq j_0$ only if $i_0,j_0$  are roots, in which case there is an edge $i_0\both j_0$. 
   The set of a-paths from $i$ to $j$ is denoted  by $\Gamma_{ij}$.
   
   The label $p_\gamma$ of an a-path $\gamma \in \Gamma_{ij}$ is the product of the labels of its edges,
\begin{equation*}
p_\gamma := \prod\nolimits_{e \in \gamma}  \alpha_e,
\end{equation*}
where $e\in\gamma$ means $e$ is an edge in the sequence $\gamma$. 
\end{definition}

An a-path contains at most two roots. We might think of three types of a-paths between two nodes: (a) those where all directed edges have direction $\from$ ($k>0,m=0$), (b) those where all directed edges have direction $\to$ ($k=0,m>0$),  and (c) those with both directions ($k,m>0$). If both nodes are roots,  then there is only one a-path consisting of an undirected edge ($k=m=0$). There cannot be both type (a) and type (b) a-paths between two nodes  as this would create a directed loop. Retuning to the genetic context, an a-path between two populations implies that one of the populations is ancestral to the other (a,b) or that they share a common ancestor (c). Due to the nature of the graph, there can be two types of a-paths between  two populations, as in Figure \ref{Fig1}(B).

  The set $\Gamma_{ii}$ contains only the empty sequence with label one.  An a-path $\gamma \in \Gamma_{ij}$, $i\not=j$, is not symmetric, meaning that it is not considered the same as the a-path $\gamma' \in \Gamma_{ji}$ composed by the edges of $\gamma$ in the opposite order. Therefore  $\Gamma_{ij} \neq \Gamma_{ji}$.  However, the labels of $\gamma$ and $\gamma'$ are identical. 
  
  For convenience, we often write an a-path as an ordered sequence of nodes  $(i_k,\ldots, i_0,j_0,\ldots,j_{m})$, leaving out $j_0$ if $i_0=j_0$.  If $\gamma=(i,\ldots,k,\ldots,j)\in\Gamma_{ij}$, then the subsequences $\gamma_1=(i,\ldots,k)\in\Gamma_{ik}$ and $\gamma_2=(k,\ldots,j)\in\Gamma_{kj}$ are a-paths as well.

 An a-path $\gamma$ is not a path in  standard graph terminology \cite{harary1969graph}.

\begin{proposition}\label{prop:sumTo1}
Consider two nodes $i,j\in\mV$ of an admixture graph $\mG$. Then $\Gamma_{ij} \neq \emptyset$. Further, the sum of the labels over $\Gamma_{ij}$ is one, $\sum_{\gamma\in\Gamma_{ij}}p_\gamma=1$. 
\end{proposition}

A tree is a connected subgraph of an admixture graph with only directed edges and at most one incoming edge to each node.  In what follows we characterize - in terms of a-paths and labels - when an admixture graph is a tree or a forest. Here a forest is a set of trees with roots  connected by undirected edges.

\begin{theorem}
\label{th:charactTree}
For an admixture graph $\mG$, the following statements are equivalent:
\begin{enumerate}
\item[(i)] for any pair of nodes, there is only one a-path  connecting them,
\item[(ii)] every a-path   has probability one,
\item[(iii)] the admixture graph consists of a forest of $R$ trees, where $R$ is the number of roots.
\end{enumerate}
\end{theorem}

\begin{definition}
Let $\ell\in\mA$ be an admixed node and $r\in\mR$ a root of an admixture graph $\mG$. Let $\Omega_{\ell r} \subseteq \Gamma_{\ell r}$  be the set of a-paths from $\ell$ to $r$ that do not contain another root. The \emph{root weight} of  $r$ with respect to $\ell$ is the probability 
\begin{equation*}
q_{\ell r} = \sum\nolimits_{\gamma \in \Omega_{\ell r}} p_\gamma. 
\end{equation*}
\end{definition}

\begin{proposition}
\label{prop:makeupDist}
Given an admixture graph, the root weights  with respect to an admixed node  form a probability distribution.
\end{proposition}

The probability $q_{\ell r}$ is the proportion of the ancestry of the node $\ell$ stemming from the root $r$.
If the admixture graph is  a forest, then  $q_{\ell r}$ is one if the node $\ell$ is in the tree with root $r$, and otherwise $q_{\ell r}=0$.

\begin{definition}\label{def:spannedgraph}
Let $\mG=(\mV,\mE,\mL)$ be an admixture graph and let $C\subseteq\mV$. We define the \emph{admixture graph spanned by $C$} as the graph  $\mG_C = \big ( \mV_C,\mE_C,\mL_C \big )$, where
\begin{equation*}
\begin{split}
\mV_C &= \big \{ i \;|\; i \text{ is in an a-path of $\Gamma_{jk}$ for some $j,k \in C$} \big \}, \\
\mE_C &= \big \{ e \;|\; e\in \mE \text{ connects two nodes of $\mV_C$ } \big \},
\end{split}
\end{equation*}
and $\mL_C$ is the set of labels  inherited from  $\mathcal G$. In particular, $\mathcal G_\mV=\mathcal G$.
\end{definition}

It is immediate to verify that the graph $\mG_C$ is an admixture graph.

\begin{proposition}\label{prop:charactLeaves}
Let $\mG$ be an admixture graph, $\mA_0\subseteq \mV$ the leaves and  $\mG_{\mA_0}$  the admixture graph spanned by $\mA_0$. Furthermore, assume every root has a child.  One of the following two equivalent conditions holds
\begin{enumerate}
\item[(i)] for each node  $k \in \mV \backslash \mA_0$, 
there is a pair of nodes $i,j \in \mA_0$ and two a-paths $\gamma\in\Gamma_{ik},\,\delta \in \Gamma_{kj}$,  such that $\gamma$ and $\delta$ only have  $k$ in common, 
\item[(ii)] for each node $k \in \mV \backslash \mA_0$, there are two nodes $i,j \in \mA_0$ and an a-path from $i$ to $j$ that includes $k$,
\end{enumerate}
if and only if $\mG=\mG_{\mA_0}$. Moreover $\mA_0$ is the smallest set spanning the graph $\mG$, in the sense that any other set that spans $\mG$ contains $\mA_0$.   
\end{proposition}

An a-path of $\mG$ between two nodes of $C$ is also an a-path of $\mG_C$ by definition.

\section{Stochastic admixture graphs}

We will assume  an admixture graph expresses  conditional independencies of a   random vector.
Specifically, an admixture graph is a chain graph (a graph with directed and undirected edges and no directed cycles), which gives rise to a special type of Markov graphical models, called chain graph models  \cite{LauritzenCausal,Frydenberg,Whittaker1990}, see Appendix A. Many models in population genetics and phylogenetics are assumed to fulfil the conditional independences expressed by a tree, which is a special type of chain graph. Here we make the natural extension to  admixture graphs. 
 
\begin{figure}[!htbp]
\centering
\includegraphics[width=10cm]{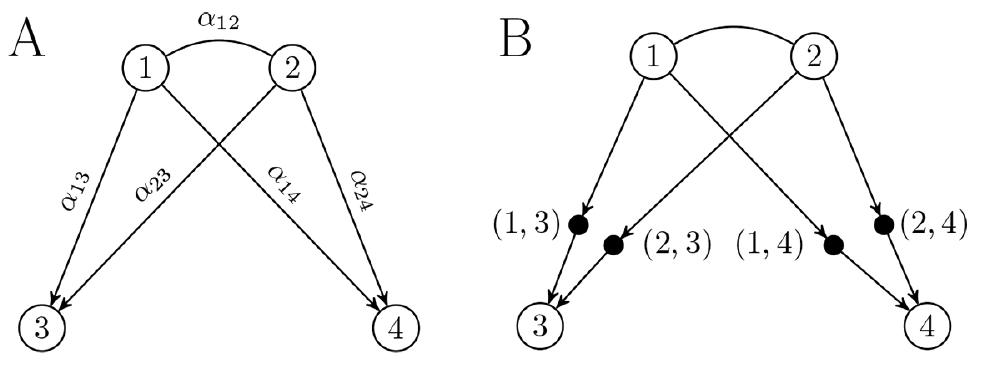}
\caption{{\bf Contribution nodes.} An example of an admixture graph (A) and its augmented graph (B). Black dots represent nodes associated with contribution variables.}\label{Fig2}
\end{figure}

Given an admixture graph $\mG$, we define the augmented  graph $\mG^*=(\mV^*,\mE^*)$ by 
$$\mV^*=\mV\cup \big\{ (i,j) \,|\, i\to j \in\mE\big\},$$
$$\mE^*=\{i\both j\,|\,i\both j\in\mE\}\cup \{ i\to(i,j)\,|\,i\to j\in\mE\}\cup \{ (i,j)\to j\,|\,i\to j\in\mE\},$$
see Figure~\ref{Fig2}. Note that $\mG^*$ is an admixture graph except for the labelling  and that an a-path of $\mG$ between two nodes $i,j\in\mV$ corresponds to a unique a-path of $\mG^*$ between the same two nodes $i,j\in\mV\subseteq\mV^*$, and vice versa.

Here and elsewhere, an equality between two random variables means equality almost surely with respect to the underlying probability measure. If $X$ is a random variable with finite expectation, then $E(X|Y)$ denotes the conditional expectation of $X$ given the random variable $Y$.

\begin{definition}
\label{def:chain}
Let $\mG$ be an  admixture graph and  $(V_i\mid i\in \mV)$ a random vector with finite expectation.  
Then, $(V_i\mid i\in \mV)$ is   a \emph{stochastic admixture graph}  over $\mG$ if there exists a random vector $(C_{ij}\mid i\to j\in\mE)$, defined on the same space as $(V_i\mid i\in \mV)$, 
with finite expectation and such that
\begin{enumerate}[label=(5.\arabic*),ref=(5.\arabic*)]
\item[(i)] $(V_j,C_{ij}\,|\,  i\in \!\pa(j), j\in\!\mV)$ is a chain graph model over $\mG^*$, 
\item[(ii)] $V_j = \sum\nolimits_{i \in \pa(j)} \alpha_{ij}C_{ij}$, $j\in \mA$, 
\item[(iii)] $E(C_{ij} | V_i) = V_i$ for $i\in \pa(j)$, $j\in \mA$. 
\end{enumerate}
The  variables $C_{ij}$, $(i,j)\in\mV^*$, are called contribution variables.
\end{definition}

The requirement that expectations are  finite ensures the existence of   conditional expectations  \cite{Hoffmann}. In the genetic context, where the variables are allele frequencies, Definition \ref{def:chain}(i) is a mathematical formalisation of the statement that evolution and gene flow happens independently along distinct lineages. Definition \ref{def:chain}(ii) states that the allele frequency of a population  is a weighted sum of the allele frequencies of the admixted populations. These frequencies are themselves the result of neutral evolution in the sense that their expectation is constant over time, Definition \ref{def:chain}(iii).

\begin{example}\label{ex:gaussian}
Assume $(V_i\mid i\in \mV)$ is a \emph{Gaussian graphical model} over a directed acyclic graph (DAG) $\mG$ \cite{Whittaker1990}. The conditional distribution of $V_j$ given the parent variables, $(V_i\mid i\in\pa(j))$, has a  Gaussian distribution with form 
\begin{equation*}\label{eq:struct}
\sum_{i\in\pa(j)} \alpha_{ij} V_i+\epsilon_j,
\end{equation*}
 where $\alpha_{ij}\not=0$ and the error terms $\epsilon_j$ are independent with expectation zero  and variance $\sigma_j^2>0$.  If $\alpha_{ij}>0$ and $\sum_{i\in\pa(j)} \alpha_{ij}=1$, then  $\epsilon_j$ can be realised as a weighted sum of independent terms  $\epsilon_j=\sum_{i\in\pa(j)} \alpha_{ij}\epsilon_{ij}$, where $\epsilon_{ij}$ has expectation zero and variance $\sigma_j^2/\alpha_{ij}$.
    By defining $C_{ij}=V_i+\epsilon_{ij}$, the three conditions of Definition \ref{def:chain} can be verified. If an arbitrary Gaussian distribution are specified for the variables of the roots of $\mG$, then   $(V_i\mid i\in \mV)$ is  a stochastic admixture graph over $\mG$ (assuming the roots are connected by undirected edges).
    
A similar remark could be said about    \emph{Structural equation models} (SEMs), where the distribution of $(V_i\mid i\in \mV)$ is allowed to take a general form while preserving \eqref{eq:struct}.
    \end{example}

\begin{example}\label{ex:wf}
The Wright-Fisher model with two alleles, say $A$ and $B$, assumes  a population of constant size $N$. The frequency of the $A$ allele evolves according to random sampling with replacement (without mutation)
$$P(X_{n+1}=\ell\mid X_n=k)=\binom{N}{\ell}\left(\frac{k}{N}\right)^{\!\ell}\left(1-\frac{k}{N}\right)^{\!N-\ell},$$
where $X_n$ denotes the number of $A$ alleles in generation $n$ \cite{Ewens1979}.  The conditional expectation of $X_{n+m}$ given $X_n$ is $E(X_{n+m}\mid X_n)=X_n$. 

Assume evolution occurs on the edges of an admixture graph according to the Wright-Fisher model and that the nodes $i$ and $(i,j)$ in  the augmented graph  represent a population at different time points. Let $V_i$ and $C_{ij}$ be the relative frequencies of the $A$ allele at these time points, and assume the population at node $j$ is the admixture of its parent populations with respective frequencies $\alpha_{ij}$, $i\in par(j)$. Then Definition \ref{def:chain}(ii)-\ref{def:chain}(iii) hold.

 The same conclusion can be made if mutation between $A$ and $B$ is allowed, and in the diffusion limit  as $N\to\infty$.  We will return to this model in Section \ref{sec:wf}.
 \end{example}

Definition \ref{def:chain}(iii)  does not in general hold for  variables associated to  the roots.

\begin{theorem}\label{th:sameRoots}
Let $\mG$ be a stochastic admixture graph, and assume $V_1,\dots,V_k$ are the variables associated with the roots. Then $E(V_i\,|\,V_j)=V_j$ holds for any pair of roots if and only if $V_1=V_2=\dots=V_k$.  
\end{theorem}

 For reasons of exposition, we assume an order on the  set of the roots, $(\mR,\prec)$, of an admixture graph.  This provides a natural way of writing undirected edges with the smallest node first, $k\both\ell$, if  $k\prec \ell$ and $k,\ell\in\mR$.   

 For two nodes $i ,j\in \mV$,  the \emph{drift} from $i$ to $j$ is  defined as
\begin{equation*}
D_{ij} := V_j-V_i.
\end{equation*} 
Note that $D_{ji}=-D_{ij}$.  Similarly, we define the \emph{partial drift} of an edge $e=k\to \ell\in\mE$ as 
\begin{equation*}
d_e = d_{k \ell } := C_{k \ell}-V_k,
\end{equation*}  
and of an edge $e=k\both \ell\in\mE$, $k\prec \ell$, between two roots as
\begin{equation*}
d_e = d_{k \ell } := V_\ell-V_k.
\end{equation*}  
 If $k\both \ell$ is an undirected edge or if $k$ is the only parent of $\ell$, then the partial drift $d_{k\ell}$  coincides with the drift $D_{k\ell}$ from $k$ and $\ell$.

Let $e=k\to\ell$ or $k\both\ell$, $k\prec\ell$, be an edge in an a-path $\gamma$. The sign  of $e$ with respect to $\gamma$ is defined as
$$ \sgn_{\gamma}(e)=\left\{\begin{array}{cl} \sgn_{\gamma}(e)=+1 & \quad\text{if}\quad\gamma=(\ldots k,\,\ell \ldots), \\ \sgn_{\gamma}(e)=-1 & \quad\text{if}\quad\gamma=(\ldots \ell,\,k \ldots), \\
\sgn_{\gamma}(e)=\phantom{+}0 & \quad\text{if}\quad e\not\in\gamma. \end{array}\right.$$

\begin{theorem}
\label{th:driftDec}
Let $(V_i\,|\, i\in \mV)$ be a stochastic admixture graph  over  $\mG$ and let $i,j \in \mV$. The drift $D_{ij}$ from $i$ to $j$  decomposes as
\begin{equation}\label{eq:driftTh}
D_{ij} = \sum\nolimits_{\gamma \in \Gamma_{ij}} \Big ( p_\gamma \sum\nolimits_{e \in \gamma} d^\gamma_e\Big ),
\end{equation}
where $d_e^\gamma = \text{sign}_\gamma(e) d_e$ is the signed partial drift with respect to $\gamma$.
\end{theorem}

If $\gamma_1\in\Gamma_{ij}$ is an a-path from $i$ to $j$, and  $\gamma_2\in\Gamma_{ji}$ is the `reversed' a-path from $j$ to $i$, then $d_e^{\gamma_1}=-d_e^{\gamma_2}$.
The signed partial drift of an undirected edge along an a-path   is independent of the order defined on $\mR$.

\begin{lemma}
\label{lem:ortoDrift}
Let $(V_i\,|\, i\in \mV)$ be a stochastic  admixture graph  over  $\mG$ and let $e_1,e_2\in\mE$ be two distinct edges, where at least one is directed. The product of their partial drifts is zero in expectation, 
\begin{equation}\label{eq:perpDrift}
E (  d_{e_1}  d_{e_2} ) = 0. 
\end{equation} 
Furthermore, for any edge $e=i\to j\in\mE$, we have $E(d_e^2) = \text{Var}(C_{ij}) -\text{Var}(V_i)$.
\end{lemma}

The second part of the lemma implies that $ \text{Var}(C_{ij}) \ge \text{Var}(V_i)$, which was first shown for the Wright-Fisher model \cite{Peter2015}. 

\begin{example}\label{ex:gaussian-variance}
Continuing with Example \ref{ex:gaussian}, for an edge $k\to \ell$, the expectation of the squared partial drift is $E(d_{k\ell}^2)=\text{Var}(V_k+\epsilon_{k\ell})-\text{Var}(V_k)=\sigma_k^2/\alpha_{k\ell}$.
\end{example}

\section{$F$-statistics}

In this section we discuss the $F$-statistics $F_2$, $F_3$ and $F_4$ \cite{Patterson2012,Reich2009,Peter2015}, and give various results for these. We assume $(V_i\mid i\in\mV)$ has finite second moments.

\begin{definition}
Let $(V_i\,|\, i\in \mV)$ be a stochastic  admixture graph  over  $\mG$ and $i,j\in\mV$. The \emph{$F_2$-statistic} between $i$ and $j$ is 
\begin{equation}\label{eq:F2basic}
F_2(i,j) = E( D_{ij}^2 ).
\end{equation}
\end{definition}

The $F_2$-statistic is non-negative and symmetric by definition.   Using Theorem \ref{th:driftDec} and Lemma \ref{lem:ortoDrift}, we might rewrite \eqref{eq:F2basic} in terms of squared partial drifts along the a-paths of $\Gamma_{ij}$, see below.

For two edges $e_1,e_2\in\mE$ and four nodes $i,j,k,\ell \in\mV$ of an admixture graph,  define the \emph{$B$-coefficient} of $(e_1,e_2)$ with respect to the pairs $(i,j),(k,\ell)$ as
\begin{align}
B_{e_1,e_2}^{(i,j)(k,\ell)} &= \sum\limits_{(\gamma_1,\gamma_2) \in \Gamma_{ij}^{e_1}\times\Gamma_{k\ell}^{e_2}} \sgn_{\gamma_1}\!(e_1)\,\sgn_{\gamma_2}\!(e_2) p_{\gamma_1}p_{\gamma_2} \nonumber \\ 
&=  \left( \sum\nolimits_{\gamma_1 \in \Gamma_{ij}^{e_1}}\sgn_{\gamma_1}\!(e)p_{\gamma_1} \right) \left( \sum\nolimits_{\gamma_2 \in \Gamma_{k\ell}^{e_2}}\sgn_{\gamma_2}\!(e)p_{\gamma_2} \right) \label{eq:Bcoef}.
\end{align}
For an edge $e\in\mE$, the \emph{$A$-coefficient} of $e$ with respect to  the pairs $(i,j),(k,\ell)$ is defined as
\begin{align}\label{eq:Acoef}
A_{e}^{(i,j)(k,\ell)} &= B_{e,e}^{(i,j)(k,\ell)}.
\end{align}
For convenience, we write  $A_{e}^{(i,j)}$ and $B_{e_1,e_2}^{(i,j)}$ if $(k,\ell)=(i,j)$.
In that case the coefficients are symmetric in $i,j$.

\begin{lemma}
\label{lem:A-B}
Let $\mG$ be an admixture graph    and $i,j,k,\ell \in\mV$. For  two edges $e_1,e_2\in\mE$, the following holds:
\begin{enumerate}[label=\textnormal{(\arabic*)}]
\item[(i)]  $-1\le B^{(i,j)(k,\ell)}_{e_1,e_2}\le 1$,
\item[(ii)] $B_{e_1,e_2}^{(i,j)(k,\ell)}=\pm 1$ if and only if $\Gamma_{ij}^{e_1}=\Gamma_{ij}$, $\Gamma_{k\ell}^{e_2}=\Gamma_{k\ell}$, $\sgn_{\gamma_1}\!(e_1)$ is independent of $\gamma_1\in\Gamma_{ij}$ and  $\sgn_{\gamma_2}\!(e_2)$ is independent of $\gamma_2\in\Gamma_{k\ell}$,
\item[(iii)] $B_{e_1,e_2}^{(i,j)(k,\ell)}= 0$ for all positive values of $\alpha_e,\, e \in \mE$, such that \eqref{eq:sum_of_alphas} is fulfilled, if and only if $\Gamma_{ij}^{e_1}=\emptyset$ or $\Gamma_{k\ell}^{e_2}=\emptyset$. 
\end{enumerate}
\end{lemma}

The lemma implies that $0 \le A_{e}^{(i,j)}\le 1$ for any edge and  $-1< B^{(i,j)}_{e_1,e_2} \leq 1$ for two undirected edges. The latter follows from (ii) and the fact that an a-path cannot pass through two distinct undirected edges.

Let $\mE_{ij}\subseteq\mE$ be the set of edges that appear in at least one path of $\Gamma_{ij}$. 
Further, let $\mE_{ij}^u\subseteq\mE_{ij}$ be the set of undirected edges and $\mE_{ij}^d\subseteq\mE_{ij}$ the set of directed edges.

\begin{theorem}
\label{th:f2Dec}
Let $(V_i\,|\, i\in \mV)$ be a stochastic  admixture graph  over  $\mG$  and $i,j\in\mV$. The $F_2$-statistic $F_2(i,j)$ decomposes as
\begin{equation}\label{eq:F2final1}
F_2(i,j)= \sum\limits_{e \in \mE^d_{ij}}  A^{(i,j)}_{e} \,E(d_{e}^2) + \sum\limits_{(e_1,e_2) \in \mE^u_{ij} \times\mE^u_{ij}}  B^{(i,j)}_{e_1,e_2} \,E(d_{e_1}d_{e_2})
\end{equation}
\end{theorem}

If $\mG_{\{i,j\}}$ has at most two roots, \eqref{eq:F2final1} might be written compactly as 
$$F_2(i,j)= \sum\limits_{e \in \mE_{ij}}  A^{(i,j)}_{e} \,E(d_{e}^2).$$
If $\mG$ is a Gaussian admixture graph, then $A^{(i,j)}_{e} \,E(d_{e}^2)=\sigma^2_kA^{(i,j)}_{e}/\alpha_{k\ell}$ is a polynomial in the labels, that is, $\alpha_{k\ell}$ cancels out, see Example \ref{ex:gaussian-variance}.

We next characterise the additivity of the $F_2$-statistic  \cite{Nei1987,Reich2009,Patterson2012}.

\begin{proposition}
\label{prop:f2Add}
Consider a stochastic admixture graph $(V_i\mid i\in\mV)$ over $\mG$, and let $i,j,k\in\mV$. If all a-paths of $\Gamma_{ij}$ pass through  $k$, then $ F_2(i,j) = F_2(i,k) + F_2(k,j) $.
 
 Assume $\mG_{\{i,j\}}$ has at most two roots.  If $ F_2(i,j) = F_2(i,k) + F_2(k,j)$ for all positive values of $E(d_e^2)$ and $\alpha_e$, $e \in \mE$, such that \eqref{eq:sum_of_alphas} is fulfilled, then all  a-paths of $\Gamma_{ij}$ pass through $k$.
\end{proposition}

If additivity holds for any node on the a-paths between any two nodes, then and only then is the admixture graph a forest (Theorem \ref{th:charactTree}).

\begin{example}
Consider the admixture graph in Figure \ref{Fig1}B. The only a-path of $\Gamma_{57}$ goes through node 2, so $F_2(5,7)=F_2(5,2)+F_2(2,7)$. For node 6 we have $F_2(5,7)=F_2(5,6)+F_2(6,7) -2F_3(6;5,7)$, but $F_3(6;5,7)$ is only zero for certain values of the parameters. In fact,
\begin{align*} 
E(D_{65}D_{67}) &= \alpha_{36}^2 E(d_{36}^2) + \alpha_{46}^2 E(d_{46}^2) - 2
\alpha_{36}\alpha_{46}E(d_{23}^2 + d_{24}^2 ),
 \end{align*}  
 which is zero for certain choices of parameters. 
\end{example}

The $F_3$- and $F_4$-statistics are defined analogously to the $F_2$-statistic
\begin{equation*}
F_3(i;j,k) = E ( D_{ij}D_{ik} ) \quad\text{and}\quad F_4(i,j;k,l) = E ( D_{ij}D_{k\ell} ),
\end{equation*}
for four nodes $i,j,k,\ell\in\mV$. Similarly to the decomposition of the $F_2$-statistic, the $F_3$- and $F_4$-statistics decompose as 
\begin{align}
&F_3(i;j,k) = \sum\limits_{e \in \mE^d_{ij} \cap \mE_{ik}^d} \!\!\!A_e^{(i,j)(i,k)} E(d_e^2) + \sum\limits_{(e_1,e_2) \in \mE^u_{ij} \times \mE_{ik}^u} \!\!\!B_{(e_1,e_2)}^{(i,j)(i,k)} E(d_{e_1}d_{e_2}) \label{eq:f3Dec} \\
& F_4(i,j;k,\ell) = \sum\limits_{e \in \mE^d_{ij} \cap \mE_{k\ell}^d} \!\!\!A_e^{(i,j)(k,\ell)} E(d_e^2) + \sum\limits_{(e_1,e_2) \in \mE^u_{ij} \times \mE_{k\ell}^u} \!\!\!B_{(e_1,e_2)}^{(i,j)(k,\ell)} E(d_{e_1}d_{e_2}). \label{eq:f4Dec}
\end{align}
(the proofs are similar to the proof for the $F_2$-statistic).

A visual method to decompose the $F$-statistics is introduced in  \cite{Patterson2012} and \cite{Reich2009}. This is formally motivated here as a consequence of the decompositions. The steps to calculate the $F_2$-statistic between two nodes $i,j\in\mV$  are the following, see Example \ref{ex:graphMethod}:

\begin{enumerate}
\item Consider all  pairs  $(\gamma_1,\gamma_2) \in \Gamma_{ij} \times \Gamma_{ij}$, including coincident pairs,
\item For each pair $(\gamma_1,\gamma_2)$, multiply  the sum of the squared partial drifts of the directed edges that are in both a-paths by $p_{\gamma_1}p_{\gamma_2}$,
\item For each pair $(\gamma_1,\gamma_2)$, sum the products of partial drifts $d_{e_1}^{\gamma_1}d_{e_2}^{\gamma_2}$ over the undirected edges $e_1 \in \gamma_1$ and $e_2 \in \gamma_2$, and multiply by $p_{\gamma_1}p_{\gamma_2}$,
\item Sum all terms above and take the expectation of the sum.
\end{enumerate}

An analogous procedure is followed to calculate the $F_3$- and $F_4$-statistics, by considering the pairs of a-paths $(\gamma_1,\gamma_2)$ between the pairs of nodes $(i,j),(i,k)$ and $(i,j),(k,\ell)$ for $F_3(i;j,k)$ and $F_4(i,j;k,\ell)$, respectively.

\begin{figure}[!htbp]
\centering
\includegraphics[width=10cm]{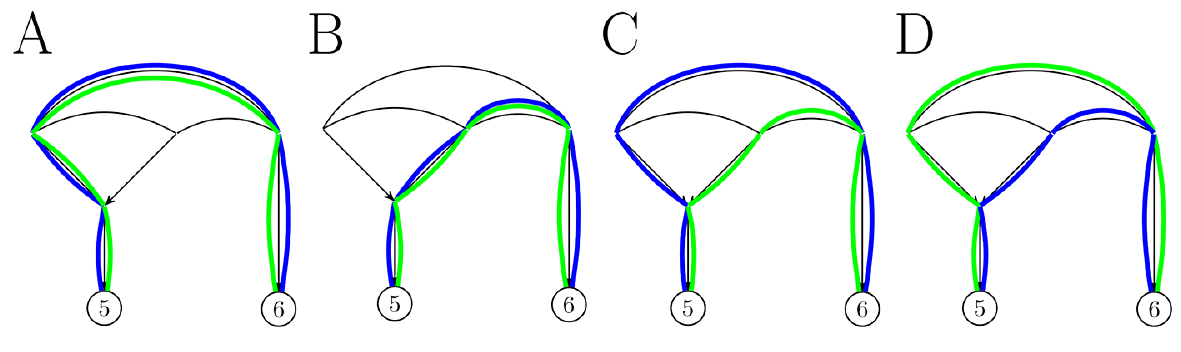}
\caption{ {\bf Visual method for the $F$-statistic.} Illustration of the visual method to calculate $F_2(5,6)$ in the admixture graphs of Figure \ref{Fig1}A. A pair of paths is represented by a blue and a green lines connecting adjacent nodes. (A-B) Pairs of coincident admixture paths sharing all their edges. (C-D) A pair of different paths where the shared directed edges are $4\to 5$, $3 \to 6$. The product of partial drifts of the two undirected edges on the blue and green paths appears in $F_2(5,6)$ (see step 3 of the visual method).}\label{Fig4}
\end{figure}

\begin{example}\label{ex:graphMethod}
Consider  $F_2(5,6)$ in the admixture graph of Figure \ref{Fig1}A. There are  two a-paths in $\Gamma_{56}$, namely $\gamma_1=(5,4,1,3,6)$ and $\gamma_2=(5,4,2,3,6)$,  resulting in four pairs of a-paths, see  Figure \ref{Fig4}A.  Applying the visual method yields:
\begin{equation*}\begin{split}
F_2(5,6)= E \Big (&  p_{\gamma_1}^2(d_{45}^2 + d_{14}^2 + d_{13}^2 + d_{36}^2) + p_{\gamma_2}^2(d_{45}^2 + d_{24}^2 + d_{23}^2 + d_{36}^2)  \\ &+2p_{\gamma_1}p_{\gamma_2}(d_{45}^2+d_{36}^2+d_{13}d_{23}) \Big ).
\end{split}\end{equation*}
By collecting terms with the same partial drifts, we obtain 
\begin{equation*}\begin{split}
F_2(5,6)&= E( d_{45}^2)+E(d_{36}^2)+ p_{\gamma_1}^2 E(d_{14}^2)  + p_{\gamma_1}^2 E(d_{13}^2)  + p_{\gamma_2}^2E(d_{24}^2)  \\ &+p_{\gamma_2}^2 E(d_{23}^2) + 2p_{\gamma_1}p_{\gamma_2}E(d_{13}d_{23}).
\end{split}\end{equation*}
We  recognize the $A$- and $B$-coefficients,
\begin{align*}
&A^{(5,6)}_{4 \to 5}=A^{(5,6)}_{3 \to 6}=1, \quad A^{(5,6)}_{1 \to 4}=A^{(5,6)}_{2 \to 4}=p_{\gamma_1}^2,  \\ &B^{(5,6)}_{2 \both 3,2 \both 3}=p_{\gamma_2}^2, \quad\quad B^{(5,6)}_{1 \both 3,1 \both 3}=p_{\gamma_1}^2, \quad\quad B^{(5,6)}_{1 \both 3,2 \both 3}=B_{2 \both 3,1 \both 3}=p_{\gamma_1}p_{\gamma_2}.
\end{align*}
\end{example}

The $F_3$-statistics can be given in terms of the $F_2$-statistics,
\begin{align*}
F_2(i,j)  = F_2(i,k)+F_2(k,j)-2F_3(k;i,j)
\end{align*}
\cite{Reich2009}.
This  shows that the $F_2$-statistic is a metric on $\mV$ (that is, it fulfils the triangular inequality) if and only if $F_3(k;i,j)\ge 0$, for all $i,j,k \in \mV$. 

If there are at most two roots then  there is at most one undirected edge and $F_3(i;j,k)$ can be written as a sum involving only $A$-coefficients, see  \eqref{eq:Acoef} and \eqref{eq:f3Dec}. Express $F_3(i;j,k)$ as the sum $F_{3+}(i;j,k) + F_{3-}(i;j,k)$, where
\begin{align*}
&F_{3+}(i;j,k) = \sum\limits_{e \in \mE_{ij} \cap \mE_{ik}} A_{e+}^{(i,j)(i,k)} E(d_e^2), \quad  F_{3-}(i;j,k) = \sum\limits_{e \in \mE_{ij} \cap \mE_{ik}} A_{e-}^{(i,j)(i,k)} E(d_e^2),
\end{align*}
and  $A_{e+}^{(i,j)(i,k)}$ and $A_{e-}^{(i,j)(i,k)}$ collect the terms  with same and opposite sign of $e$ in $(\gamma_1, \gamma_2) \in \Gamma_{ij}^e \times \Gamma_{ik}^e$, respectively. 

The next statement relates the  sign of the $F_3$-statistic to the topology of the admixture graph.

\begin{proposition}\label{prop:F3configs}
Let $\mG$ be an admixture graph and $C=\{i,j,k\}\subseteq\mV$  such that $\mG_C$ has at most two roots.  Then $F_{3-}(i;j,k) \neq 0$ if and only if there exist $e \in \mE_{ij} \cap \mE_{ik}$ and a pair of a-paths $(\gamma_1,\gamma_2) \in \Gamma^e_{ij} \times \Gamma^e_{ik}$ such that
\begin{align*}
\gamma_1 &= i \from \dots \to j \quad\text{ or }\quad \gamma_1 = i \from \dots \both j, \\
\gamma_2 &= i \from \dots \to k \quad\text{ or }\quad \gamma_2 = i \from \dots \both k,
\end{align*} 
and $\sgn_{\gamma_1}(e)=-\sgn_{\gamma_2}(e)$.  
\end{proposition}

\section{The Wright-Fisher model}
\label{sec:wf}

In this section we consider the Wright-Fisher model in the diffusion limit as the population size  becomes infinite \cite{Ewens1979}, see also Example \ref{ex:wf}. 

\begin{definition}
A \emph{Wright-Fisher admixture graph} $(V_i\vert i\in\mV)$ over $\mG$ is a stochastic admixture graph such that $V_i$, $i\in V$, are the frequencies of a particular allele in the populations represented by  the nodes of the graph, and such that $C_{ij}$ conditional on $V_i=v\in[0,1]$, $i\in\pa(j)$, $j\in\mV$, is  distributed as the frequency in a Wright-Fisher diffusion process  at time $\tau_{ij}$, given it starts at frequency $v$, and $\tau_{ij}$ is the population scaled time between  the nodes $i$ and $(i,j)$.
\end{definition}

\cite[chapter 13]{Nei1987} showed that the expected heterozygosity  of an allele declines exponentially over time, 
\begin{equation*}
E(V_1(1-V_1))=(1-e^{-\tau_{12}})E(V_2(1-V_2)),
\end{equation*}
where $V_1$ is the frequency at time zero and $V_2$ the frequency $\tau_{12}$  population scaled time units later. We generalise this formula to a Wright-Fisher admixture graph.

\begin{proposition}\label{prop:heteroWF}
Consider a Wright-Fisher admixture graph $(V_i\mid i\in\mV)$ over $\mG$.  Then the expected heterozygosities fulfil
\begin{align*}
E(V_j(1-V_j)) &= \sum_{(r_1,r_2)\in \mR\times\mR}  D_{r_1,r_2}^j E(V_{r_1}(1-V_{r_2})),  \quad j\in\mV,
\end{align*}
where $D^j_{r_1,r_2}$ are non-negative constants depending on the edge labels and the population scaled times. Furthermore,
$$\sum_{r_1,r_2} D^j_{r_1,r_2}\le 1,  \quad j\in\mV.$$
\end{proposition}

If all population scaled times $\tau_{ij}$, $i\to j\in\mE$, between two nodes are set to zero, then the sum of $D^j_{r_1,r_2}$ over $r_1,r_2$ is precisely one (see the proof of the. proposition). In that case, $E(V_j(1-V_j))$ is the probability of drawing two different alleles from the root populations. With probability $D_{r_1,r_2}^j$, alleles from population $r_1$ and $r_2$ are drawn. In general, when $D_{r_1,r_2}^j$ does not sum to one, heterozygosity is lost.

If there is only one root, then the sum consists of just one term, hence all expected heterozygosities can be expressed in terms of the expected heterozygosity of the root variable.  
   \cite{Nei1987} showed that the partial drift of an edge $e=i\to j$ fulfils 
   \begin{equation}\label{eq:de}
   E(d_e^2)=(1-e^{-\tau_{ij}})E(V_i(1-V_i)),
   \end{equation}
    so that the $F_2$-statistics can be expressed in terms of the expected heterozygosities of the root variables (Theorem \ref{th:f2Dec}).

To end we consider a few examples. The linear Wright-Fisher admixture graph with edges  $1\to 2$, $2\to 3$ was already studied in \cite[chapter 13]{Nei1987} and it was found that
$$F_2(2,3)=e^{-\tau_{12}}(1-e^{-\tau_{23}})E(V_1(1-V_1)).$$
It is also a consequence of \eqref{eq:de} and additivity of the $F_2$-statistic.

 In a Wright-Fisher admixture graph with edges   $1\to 2$, $2\to 3$, $2\to 4$, $3\to 5$,  $4\to 5$, such that there are two overlapping a-paths from $1$ to $5$, we have 
\begin{align*} 
F_2(1,5) &= \big ( 1 - \alpha_{35}^2e^{-\hat\tau_1}-\alpha_{45}^2 e^{-\hat\tau_2}- 2\alpha_{35}\alpha_{45}e^{-\tau_{12}} \big )E(V_1(1-V_1)),
 \end{align*}
where $\hat\tau_1=\tau_{12}+\tau_{23}+\tau_{35}$ and  $\hat\tau_2=\tau_{12}+\tau_{24}+\tau_{45}$.

\section*{Acknowledgement}

CW acknowledges support from the Danish Research Council.


%

\section*{Appendix A}

Consider a graph $\mG=(\mV, \mE)$, where $\mV,\mE$, are the nodes and the edges, respectively, and edges are directed or undirected.  A directed edge from $i$ to $j$ and an undirected edge between $i$ and $j$ are written as $i \to j$ and $i \both j$, respectively. A \emph{path} from node $i$ to node $j$ is a sequence of adjacent nodes $i_0,i_1,......,i_K$, where $i_0=i,i_K=j$, and either $i_k \to i_{k+1}$ or $i_k \both i_{k+1}$, for $k=0,\dots,K-1$.  A \emph{cycle} is a path with $i=j$. A cycle is directed if $i_k \to i_{k+1}$ for  $k=0,\dots,K-1$. (A path is \emph{not} the same as an a-path.)

A graph where the nodes are connected by either directed or undirected edges and without directed cycles is called a \emph{chain graph}. 

The node set $\mV$ of a chain graph $\mG$ can be partitioned into unique \emph{blocks} $B_1,\dots,B_N$  such that  $B_1=\mR$, and if two nodes $i,j\in\mV$ are connected by a sequence of directed edges  $i \to i_1,i_1\to i_2,\ldots, i_k\to j$ in $\mE$, then $i \in B_{n_i},\, j \in B_{n_j}$ and $n_i<n_j$. Two nodes $i,j$ are said to be in the same \emph{component} if they are connected by an undirected or directed path from $i$ to $j$ and from $j$ to $i$. Note that a component does not coincide with a block. In Figure \ref{Fig1}B each node is a distinct component, but the blocks are $B_1=\{1\}$, $B_2=\{2,8\}$, $B_3=\{3,4\}$, $B_4=\{5,6,7\}$.

A subset $A \subseteq \mV$ of a chain graph $\mG$ induces a graph $\mG(A) = (A,\mE(A))$, where $\mE(A)$ contains the edges of $\mG$ whose nodes are in $A$

The \emph{border} of a subset  $\mA\subseteq\mV$  is defined as
\begin{equation*}
bd(\mA) := \big \{ i \in \mV \;:\; i \rightarrow j \text{ or } i \both j \,\text{for some $j \in A$} \big \},
\end{equation*}
and the moral graph of $\mG$ is $\mG^m=(\mV, \mE^m)$, where $\mE^m$ consists of the union of
\begin{itemize}
\item the set $\mE^u$ consisting of the edges of $\mE$ made undirected,
\item the set of undirected edges of $\mE$ that connect pairs of nodes that are in the border of a component of $\mG$.
\end{itemize}
In Figure \ref{Fig1}B the moral graph is obtained by making all edges undirected, and by adding an undirected edge between $3$ and $4$.

Let $A,B,C$ be three disjoint subsets of $\mV$. The sets $A$ and $B$ are \emph{separated} by $C$ when every path from $i\in A$ to $ j \in B$ (or vice versa) includes at least one node of $C$. Given $A \subseteq V$,  the set of ancestors of $A$, $an_G(A)$, is the subset of nodes of $\mV$ having at least one path to some node of $A$.

A random vector $(U_i| i \in \mV)$ is a chain graph model over $\mG$ if it fulfils the \emph{global $\mG$-Markovian} (GM) property,
\begin{equation*}
(U_i| i\in A) \,\,\perp\,\, (U_i|i\in B) \quad |\quad (U_i|i\in C),
\end{equation*}
whenever $C$ separates $A$ and $B$ in $\mG_{an(A \cup B \cup C)}^m$ and $X\perp Y| Z$ denotes that the random vectors $X$ and $Y$ are conditionally independent given $Z$.

Let $(V_i| i\in\mV)$ be  a stochastic admixture graph over $\mG$. The Markov structure of the chain graph implies in particular that  two contribution variables $C_{ij},C_{k \ell}$, where $i,k$ are not necessarily distinct,  are conditionally independent given the variables $V_i,V_k$ of their parents, that is,
\begin{equation}\label{eq:mpCont1}
C_{ij} \perp C_{k \ell} \mid (V_i,V_k).
\end{equation}
Further, for a contribution variable $C_{ij}$, let $B_{n_i}$ be the block in which node $i$ is located. Then
\begin{equation}\label{eq:mpCont2}
C_{ij} \perp \left(V_k\mid k \in\bigcup\nolimits_{n=1}^{n_i} B_n\right) \mid V_i.
\end{equation}
As a consequence of \eqref{eq:mpCont2}, if a node $k$ is in $\bigcup\nolimits_{n=1}^{n_i} B_n$, it follows that
\begin{equation}\label{eq:mpCont2Single}
C_{ij} \perp V_k \mid V_i.
\end{equation}

\section*{Appendix B}

Proofs of statements in the main text are presented here.  We start by listing in Table \ref{tab:path} the possible forms of an a-path between two distinct nodes $i,j\in\mV$. In a,e) there might  be only one edge, in the others at least two.
For two a-paths $\gamma,\delta$, the concatenation of $\gamma$ with $\delta$ is the sequence of edges in $\gamma$ followed by those of $\delta$. It might or might not be an a-path.

\begin{table}[h]
\centering
\begin{tabular}{lcccccccccccc}
&a) & $i$& $\from$  & $\ldots$  & $\from$ &  $j$ &d) & $i$& $\both$  & $\ldots$  & $\to$ &  $j$ \\
&b) & $i$& $\from$  & $\ldots$  & $\both$ &  $j$ &e) & $i$& $\to$  & $\ldots$  & $\to$ &  $j$ \\
&c) & $i$& $\from$  & $\ldots$  & $\to$ &  $j$ &&&&& \\
\end{tabular}
\caption{Possible admixture paths according to Definition \ref{defn:path}.}
\label{tab:path}
\end{table}

\noindent
\begin{lemma} \label{lem:path-statement}
Let $i,j\in \mV$ be two distinct nodes of an admixture graph $\mG$, such that not both of them are roots. Then  either all a-paths $\gamma\in\Gamma_{ij}$ end with a directed edge $k\to j$, for some $k\in\mV$, or all a-paths $\gamma\in\Gamma_{ij}$ start with a directed edge $i\from k$, for some $k\in\mV$, where $k$ might depend on $\gamma$. \qed
\end{lemma}

\medskip
\noindent
{\it Proof of Lemma \ref{lem:path-statement}.} It follows by inspecting the possibilities for a-paths.

\medskip
\noindent
\begin{lemma} \label{lem:path-statement2}
Let $i,j\in \mV$ be two distinct nodes of an admixture graph $\mG$, such that $i\not\in\mR$ and assume that all a-paths $\gamma\in\Gamma_{ij}$ start with an edge $i\from k$, for some $k\in\mV$ (as in Lemma \ref{lem:path-statement}). Then 
$$\Gamma_{ij}=\cup_{\ell\in\pa(i)}\{(i\from \ell,\gamma)\mid \gamma\in\Gamma_{\ell j}\},$$
where $(i\from \ell,\gamma)$ denotes the concatenation of the a-path $i\from \ell$ with $\gamma$.  \qed
\end{lemma}

\medskip
\noindent
{\it Proof of Lemma \ref{lem:path-statement2}.} It follows trivially that $\Gamma_{ij}$ is included in the right hand side of the equality. To prove the converse we need to show that $(i\from i',\gamma)$, $\gamma\in\Gamma_{i'j}$ is an a-path from $i$ to $j$, that is, the edges are directed according to Definition \ref{defn:path} (see Table \ref{tab:path}) and $i$ does not belong to $\gamma$. By inspection of the possibilities for $\gamma$, it shows that the edges have the correct directions (including undirected).  Assume $i$ is in $\gamma$. It cannot be a root by assumption. Then by inspection, either there is a subsequence $i'\from \ldots \from i$ of $\gamma$, or a subsequence $i\to \ldots \to j$. In the former case, this yields a directed cycle from $i$ to $i$ via $i'$ by adding the edge $i'\to i$. In the latter case, it yields an a-path from $i$ to $j$, violating the assumption of the starting edge. Hence the proof is completed. \qed
 
\medskip
\noindent
\begin{lemma} \label{lem:path-statement3}
Let $i,j,k\in\mV$ be distinct nodes of an admixture graph $\mG$, and assume any a-path of $\Gamma_{ij}$ passes through  $k$. Then it holds that
$$\Gamma_{ij}=\Gamma_{ik}\times\Gamma_{kj},$$
where $(\gamma,\delta)\in \Gamma_{ik}\times\Gamma_{kj}$ denotes concatenation of $\gamma$ and $\delta$.   \qed
\end{lemma}

\medskip
\noindent
 {\it Proof of Lemma \ref{lem:path-statement3}. } 
 We will prove the two sets are included into each other. It trivially holds that $\Gamma_{ij}\subseteq \Gamma_{ik}\times\Gamma_{kj}$. Next we prove the converse inclusion. According to Lemma \ref{lem:path-statement}, we might assume all a-paths $\gamma\in\Gamma_{ik}$ start with a directed edge $i\from i'$, $i'\in\pa(i)$, hence the same is true for the a-paths of $\Gamma_{ij}$. According to Table \ref{tab:path} this yields $3\cdot 5=15$ possibilities for the forms of $(\gamma,\delta)$. In the cases (a,a), (a,b), (a,c), (a,d), (a,e), (b,e), (c,e) the concatenation of $\gamma$ with $\delta$ is an a-path from $i$ to $j$. The cases (b,a), (b,b), (b,c), (c,d) give impossible constraints on $k$, hence they cannot occur.  The case (b,d) gives $i\from\ldots\from k_1\both k\both k_2\to \ldots\to j$ for two roots $k_1,k_2\in\mR$. Hence there is an a-path $i\from\ldots\from k_1\both k_2\to \ldots\to j$, not going though $k$, which is impossible. The remaining three cases are

\medskip
\begin{tabular}{lcccccccccc}
 & & & $\gamma$ & & & & $\delta$ & & & \\
(c,a) & $i$& $\from$  & $\ldots$  & $\to$ &  $k$ & $\from$  & $\ldots$  & $\from$ &  $j$ \\
(c,b) & $i$& $\from$  & $\ldots$  & $\to$ &  $k$ & $\from$  & $\ldots$  & $\both$ &  $j$ \\
(c,c) & $i$& $\from$  & $\ldots$  & $\to$ &  $k$ & $\from$  & $\ldots$  & $\to$ &  $j$ 
\end{tabular}

\medskip
\noindent
The node $k$ cannot be a root, $k\not\in\mR$. We will construct an a-path from $i$ to $j$, based on $\gamma$ and $\delta$, that does not go through $k$, hence (c,a), (c,b), (c,c) cannot occur. Specifically, there are three possibilities for the a-path $\gamma$,

\medskip
\begin{tabular}{ccccccccc}
 $i$& $\from$  & $\ldots$ & $\from$ & $i'$ & $\to$ &  $\ldots$  & $\to$ &  $k$ \\
 $i$& $\from$  & $\ldots$  & $\from$ &  $r$ & $\to$  & $\ldots$  & $\to$ &  $k$ \\
 $i$& $\from$  & $\ldots$  & $\from$ &  $r$ \,\,\,$\both$\,\,\, $r'$ & $\to$  & $\ldots$  & $\to$ &  $k$ 
\end{tabular}

\medskip
\noindent
In the first case, there is $r\in\mR$ and an a-path $r\to\ldots\to i'$ by definition of an admixture graph. It does not contain $k$ as this would create a directed cycle. Consequently, in all three cases, there is $r\in\mR$ and an a-path $r\to\ldots\to i$, not containing $k$. 

For $\delta$ we can do similarly. In the first case above, there exists $r'\in\mR$ and an a-path $r'\to\ldots\to j$, not containing $k$ as this would create a directed cycle. In the second case, $j\in\mR$ and we define $r'=j$. In the third case, we proceed as for $\gamma$ and conclude there is $r'\in\mR$ and an a-path $r'\to \ldots\to j$, not containing $k$.

Now consider the  a-path $\eta$ from $r$ to $i$, and the a-path $\eta'$ from $r'$ to $j$, constructed above ($r=r'$ or $r\not=r'$). If $\eta$ and $\eta'$ do not share any nodes, except perhaps for $r,r'$, then their concatenation is an a-path $(\eta,\eta')\in\Gamma_{ij}$, not containing $k$.  If they do share nodes, chose sub-a-paths $\zeta\colon \ell\to\ldots\to i$ and $\zeta'\colon\ell\to\ldots \to j$, sharing only the node $\ell$. Then their concatenation is an a-path $(\zeta,\zeta')\in\Gamma_{ij}$, not containing $k$. Hence, none of the cases (c,a), (c,b), (c,c) are valid, and the lemma holds.

\medskip
\noindent
{\it Proof of Proposition \ref{prop:sumTo1}.}
If $i=j$ then the statements are true by definition.
Given distinct $i,j\in\mA$, there exists $r_1,r_2\in\mR$ and two a-paths $r_1 \to i_1 \to  \ldots \to i_{k-1} \to i$ and $r_2 \to j_1 \to\ldots \to j_{k'-1} \to j$ for some nodes $i_1,\ldots,i_{k-1},  j_1,   \ldots,\\ j_{k'-1}\in\mA$, and $k,k'\ge 1$ (because the admixture graph is connected and there are no directed cycles). Either all these nodes are distinct (except perhaps for $r_1,r_2$) in which case they form an a-path from $i$ to $j$ by concatenation, or there is $i_\ell=i'_{\ell'}$ for some $\ell,\ell'$. Choose $\ell,\ell'$ such that $\ell+\ell'\le k+k'$ is as large as possible. Then $i\from i_{k-1}\from \ldots\from i_{\ell+1} \from i_\ell \to i'_{\ell'+1}\to\ldots \to i'_{k'-1} \to j$ is an a-path from $i$ to $j$ by definition of an a-path. There cannot be any repeated nodes, otherwise $\ell+\ell'$ is not as large as possible.  If $i,j\in\mR$, then they are trivially connected by an a-path. If $i\in\mR$ and $j\in\mA$, then there is an a-path $j\from i_{k-1}, \ldots\from i_1\from i$ (potentially with $i_1\both i$ and $i_1$ a root) for some $i_1,\ldots,i_{k-1}\in\mA$ (as before).  Hence $\Gamma_{ij}\not=\emptyset$.

To prove the second part of the proposition, we proceed by induction in the length of the  a-paths. 
For $i,j\in\mV$, consider $\Gamma_{ij}$ and let $l_\gamma$ denote the number of edges in an a-path $\gamma\in\Gamma_{ij}$. Assume $l_\gamma\le 0$ for all $\gamma\in\Gamma_{ij}$. Hence $i=j$ and $\sum_{\gamma\in\Gamma_{ij}}p_\gamma=1$ by definition.

Assume now the statement holds if $\sum_{\gamma\in\Gamma_{ij}}p_\gamma=1$ and $l_\gamma\le k$, $\gamma\in\Gamma_{ij}$, for some $k\ge 1$. Consider two nodes $i,j\in\mV$ such that all a-paths between them fulfil $l_\gamma\le k+1$ and start with an edge $i\from k$ for some $k\in\mV$ (otherwise exchange the roles of $i$ and $j$). Then, using   Lemma \ref{lem:path-statement2} to decompose the a-paths according to the parents of $i$, we obtain
$$\sum_{\gamma\in\Gamma_{ij}}p_\gamma=\sum_{\ell\in par(i)} \sum_{\gamma\in\Gamma_{\ell j}} \alpha_{i\ell}p_\gamma=\sum_{\ell\in par(i)} \alpha_{i\ell}\sum_{\gamma\in\Gamma_{\ell j}} p_\gamma=\sum_{\ell\in par(i)} \alpha_{i\ell}=1,$$
 as all a-paths in $\Gamma_{\ell j}$ must have length at most $k$.     \qed

\medskip
\noindent 
{\it Proof of Theorem \ref{th:charactTree}.} 
We will prove $(ii)\Rightarrow (i)\Rightarrow (iii)\Rightarrow (ii)$. Assume (ii).  If the label $p_\gamma$ of an a-path between $i$ and $j$ is one, then $\Gamma_{ij}=\{\gamma\}$, according to Proposition \ref{prop:sumTo1}. It proves (i).  Let $\mA_r$ be the nodes in $\mA$ for which there is an a-path from $r\in\mR$ to a node in $\mA$, not involving any other root. Any $i\in\mA$ is in at least one $\mA_r$, and cannot be in two such sets $\mA_{r_1},\mA_{r_2}$, because then there would be a-paths $(r_1,\ldots,i)$ and $(r_1,r_2,\ldots,i)$ from $r_1$ to $i$, contradicting uniqueness of a-paths (i). The set $\{r\}\cup\mA_r$ with the inherited edges form a  tree. It proves (iii).   The last implication is straightforward using the definition of a forest and taking into account the undirected edges between the roots, so (ii) is proven.
\qed

\medskip
\noindent
{\it Proof of Proposition \ref{prop:makeupDist}.} 
Consider $r_1 \in \mR$. Any a-path $\gamma_1\in\Gamma_{\ell r_1}$ contains  one root  or two roots. In the latter case, the a-path is the concatenation of an a-path $\gamma'\in\Omega_{\ell r_2}$, $r_2\not=r_1$, with the edge $r_2\both r_1$ and $p_{\gamma}=p_{\gamma'}$. It follows that
\begin{equation*}
\sum\limits_{r \in \mR} q_{\ell r} =\sum\limits_{r \in \mR}\sum\limits_{\gamma \in \Omega_{\ell r}} p_\gamma = \sum\limits_{\gamma \in \Gamma_{\ell r_1}} p_\gamma = 1.
\end{equation*}
\qed

\medskip
\noindent
{\it Proof of Proposition \ref{prop:charactLeaves}.} 
For the if and only if, assume (ii). By Definition \ref{def:spannedgraph}, we have $\mV_{\mA_0}=\mV$ and $\mE_{\mA_0}=\mE$. Oppositely, if  $\mG_{\mA_0}=\mG$, then again by Definition \ref{def:spannedgraph}, (ii) holds.

We next prove the equivalence between (i) and (ii). (ii) trivially implies (i).  To prove the converse,  let $\gamma\in\Gamma_{ik}$, $\delta\in\Gamma_{kj}$ be as stated, sharing only the node $k\in \mV\setminus\mA_0$ and $i,j\in\mA_0$. There are five possibilities for a-paths, see Table \ref{tab:path}. As $i,j\in\mA_0$, then d,e) are not an option for $\gamma$, and a,b) are not an option for $\delta$, yielding nine possible combinations.  By inspection,  (a,c), (a,d), (a,e), (b,e), (c,e) yield valid a-paths from $i$ and $j$, hence (ii) holds in these case. The cases (b,c), (c,d) yield impossible constraint on $k$, and thus cannot occur. The two remaining cases (b,d), (c,c) give

\medskip
\begin{tabular}{lcccccccccc}
 & & & $\gamma$ & & & & $\delta$ & & & \\
(b,d) & $i$& $\from$  & $\ldots$  & $\both$ &  $k$ & $\both$  & $\ldots$  & $\to$ &  $j$ \\
(c,c) & $i$& $\from$  & $\ldots$  & $\to$ &  $k$ & $\from$  & $\ldots$  & $\to$ &  $j$ 
\end{tabular}

\medskip
\noindent
 Since $k\not\in\mA_0$ then there is an a-path, say $\eta$, $k\to \ldots \to k'$, with $k'\in\mA_0$ (in the first case we use that every root has a child). If this a-path does not share a node with $\gamma$ (similarly, with $\delta$) except from $k$, then $\gamma$ concatenated with $\eta$ is an a-path from $i$ to $k'$ through $k$, as required. Otherwise, if $\eta$  shares nodes with both a-paths, let $k''\not=k$ (say in $\delta$, similar if in $\gamma$) be the first such node counting from $k$. Let $\eta'$ be the a-path from $k$ to $j$ given by $k\to \ldots\to k''\to \ldots\to j$ by concatenating the part of $\eta$ ending at $k''$ with the part of $\delta$ beginning at $k''$. The edge from $k''$ towards $j$ cannot be $\both$ nor $\from$ for the following reasons. In the former case, $\to k''\both$ which is impossible. In the latter case, $\delta$ takes the form $k\from\ldots\from k''\from\ldots\from j$, creating a directed loop from $k$ to $k''$ and back to $k$ via $\eta'$.  Finally,  $\gamma$ concatenated with $\eta'$ is an a-path from $i$ to $j$ through $k$, as required.

\medskip
\noindent
{\it Proof of Theorem \ref{th:sameRoots}.} 
A general reference is  \cite{Hoffmann}.
Let $V_i,V_j$ be two root variables, corresponding to distinct roots in $\mR$. By definition of conditional expectation (first equality) and by assumption (second equality)
$$\int_A V_i\,dP=\int_A E(V_i\mid V_j)\,dP=\int_A V_j\,dP$$
for all $R_j$-measurable subsets $A$. Similarly, by exchanging the role of $V_i$ and $V_j$, the same holds for all $V_i$-measurable subsets. Hence,
$$\int_A V_i\,dP=\int_A V_j\,dP$$
for all $(V_i,V_j)$-measurable subsets $A$. Consequently, using that $V_i,V_j$ have finite expectation by assumption, it follows that $V_i=V_j$.
\qed

\medskip
\noindent
{\it Proof of Theorem \ref{th:driftDec}.} We prove the statement by induction on the maximum number of edges $n$ of the a-paths $\gamma \in \Gamma_{ij}$. If $n=0$, then $i=j$ and there is nothing to prove.
Assume now \eqref{eq:driftTh} for is true for $n=\ell$ and consider $n=\ell+1$ for some $\ell \geq 1$. Further, assume that $i$ is such that any a-path of $\Gamma_{ij}$ involves a parent of $i$  as the first node (Lemma \ref{lem:path-statement}).
Then using the inductive hypothesis, Proposition \ref{prop:sumTo1},  Definition \ref{def:chain}(ii), Lemma \ref{lem:path-statement2} and \eqref{eq:sum_of_alphas}, it holds that
\begin{align*}
\sum\limits_{\gamma \in \Gamma_{ij}} p_\gamma \sum\limits_{e \in \gamma} d^\gamma_e &= \sum\limits_{i' \in par(i)} \sum\limits_{\gamma' \in \Gamma_{i'j}} \alpha_{i'i} p_{\gamma'} \Big ( d_{i'i}^{(i,i')} + \sum\limits_{e \in \gamma'} d_e^{\gamma'} \Big )\\
&= \sum\limits_{i' \in par(i)} \alpha_{i'i} \Big ( d_{i'i}^{(i,i')} + \sum\limits_{\gamma' \in \Gamma_{i'j}} p_{\gamma'} \sum\limits_{e \in \gamma'} d_e^{\gamma'} \Big )\\
& = \sum\limits_{i' \in par(i)} \alpha_{i'i} \Big ( C_{i'i} - V_{i'} + V_{i'} - V_j \Big ) \\
& = \sum\limits_{i' \in par(i)} \alpha_{i'i}C_{i'i} - \sum\limits_{i' \in par(i)} \alpha_{i'i} V_j = V_i - V_j.
\end{align*}
If any a-path of $\Gamma_{ij}$ does not involve a parent of $i$  as the first node after $i$, then it holds that any a-path involves a parent of $j$ as second last node (Lemma \ref{lem:path-statement2}).  Note that $D_{ij}=-D_{ji}$, and that the decomposition holds for $D_{ji}$. Thus
\begin{align*}
D_{ij} &= - \sum\limits_{\gamma \in \Gamma_{ji}} p_\gamma \sum\limits_{e \in \gamma} d_e^\gamma = \sum\limits_{\gamma \in \Gamma_{ij}} p_\gamma \sum\limits_{e \in \gamma} d_e^\gamma, 
\end{align*}
using that the partial drifts change sign when an a-path is reversed. 
\qed

\medskip
\noindent
{\it Proof of Lemma \ref{lem:ortoDrift}. } 
Let $e_1=i\to j$ and $e_2=k\to \ell$ (with one of the two potentially being undirected). Further, assume that the blocks $B_{n_i},B_{n_k}$, associated to the nodes $i,k$ are such that $n_i \geq n_k$. 
The expected value in \eqref{eq:perpDrift} can be rewritten as
\begin{equation*} 
E ( d_{ij}  d_{k\ell}) = E ( C_{ij} C_{k \ell} ) - E ( C_{i j} V_k ) - E ( C_{k \ell} V_i ) + E ( V_i V_k ). 
\end{equation*}
Consider the first expectation,
\begin{align}
E ( C_{ij} C_{k \ell} ) &= E( E(C_{ij} C_{k \ell} | V_i,V_k))   
	 = E( E(C_{ij} | V_i,V_k) E(C_{k\ell} | V_i,V_k)) \label{eq:calc1} \\
	 &= E( E(C_{ij} | V_i) E(C_{k\ell} | V_i,V_k)) 
	 = E(V_i  E(C_{k\ell} | V_i,V_k)). \label{eq:calc2}
\end{align}
Here we used \eqref{eq:mpCont1}, \eqref{eq:mpCont2Single} to derive  \eqref{eq:calc1}, and Definition~\ref{def:chain} to derive \eqref{eq:calc2}.  With similar considerations, it follows that
\begin{align*}
E (C_{k\ell} V_i) &= E( E( C_{k\ell}  V_i| V_i,V_k)) = E( V_i  E(C_{k\ell} | V_i,V_k)).
\end{align*}
Finally,
\begin{align}
E(C_{ij} V_k) &= E( E( C_{ij}  V_k| V_i )) \label{eq:res1} \\
	& = E( E(C_{ij} | V_i) E(V_k | V_i)) \label{eq:res2} \\
	& = E(V_i E(V_k | V_i))= E( E(V_i V_k|V_i))=E(V_i V_k),  \label{eq:res3} 
\end{align}
using \eqref{eq:mpCont2Single} and Definition~\ref{def:chain}(ii) to derive \eqref{eq:res1} and \eqref{eq:res2}, respectively, and properties of the conditional expectation to derive \eqref{eq:res3}.  Combining it all yields $E (d_{ij}  d_{k \ell} )=0$.
Finally consider $E(d_{ij}^2)=E(C_{ij}^2 + V_i^2 -2C_{ij}V_i)$. Using the definition of variance and Definition \ref{def:chain}(ii), it follows that
$$ E(C_{ij}^2) = Var(C_{ij}) + E(V_i)^2 \quad\text{and}\quad E(V_i^2) = Var(V_i) + E(V_i)^2. $$
By conditioning on $V_i$ and using Definition \ref{def:chain}(ii) again, we have $E(C_{ij}V_i)=-Var(V_i)-E(V_i)^2$. Using the linearity of the expectation it can be concluded that $E(d_{ij}^2) = Var(C_{ij}) - Var(V_i)$.
  \qed

\medskip
\noindent
{\it Proof of Theorem \ref{th:f2Dec}.} 
Rewrite the definition of $F_2(i,j)$ using the  decomposition of drift in \eqref{eq:driftTh},
\begin{equation}\label{eq:F2_1}
\begin{split}
F_2(i,j)&= E \Biggl [ \; \left(\sum\limits_{\gamma_1 \in \Gamma_{ij}}  \left(p_{\gamma_1} \sum\limits_{e_1 \in \gamma_1} d^{\gamma_1}_{e_1}\right)\right)  \left(\sum\limits_{\gamma_2 \in \Gamma_{ij}}\left( p_{\gamma_2} \sum\limits_{e_2 \in \gamma_2} d^{\gamma_2}_{e_2}\right)\right)  \; \Biggr ].
\end{split}\end{equation}
Distributing the products and exploiting the linearity of the expectation, \eqref{eq:F2_1} is equivalent to
\begin{align*}
F_2(i,j) &=\sum\limits_{(\gamma_1,\gamma_2) \in \Gamma_{ij} \times \Gamma_{ij}} \sum\limits_{e_1 \in \gamma_1}\sum\limits_{e_2 \in \gamma_2} p_{\gamma_1}p_{\gamma_2} E ( d^{\gamma_1}_{e_1}d^{\gamma_2}_{e_2} ), \\
&=  \sum\limits_{(e_1,e_2) \in \mE_{ij}\times\mE_{ij}} \; \sum\limits_{(\gamma_1,\gamma_2) \in \Gamma_{ij}^{e_1} \times \Gamma_{ij}^{e_2}}         p_{\gamma_1}p_{\gamma_2} E ( d^{\gamma_1}_{e_1}d^{\gamma_2}_{e_2} ).
\end{align*}
Observe that $\mE_{ij}$ is the disjoint union of $\mE_{ij}^d$ and $\mE_{ij}^u$. Moreover, the product of two distinct edges, where at least one is directed, has expectation zero, see Lemma \ref{lem:ortoDrift}. Thus
\begin{align*}
F_2(i,j) &= \sum\limits_{e \in \mE^d_{ij}} \; \sum\limits_{(\gamma_1,\gamma_2) \in \Gamma_{ij}^{e} \times \Gamma_{ij}^{e}}  \sgn_{\gamma_1}\!(e)\,\sgn_{\gamma_2}\!(e) p_{\gamma_1}p_{\gamma_2} \,E ( d_{e}^2 ) \nonumber 
\\ &+ \sum\limits_{(e_1,e_2) \in \mE^u_{ij} \times\mE^u_{ij}} \; \sum\limits_{(\gamma_1,\gamma_2)  \in \Gamma_{ij}^{e_1} \times \Gamma_{ij}^{e_2}} \sgn_{\gamma_1}\!(e_1)\,\sgn_{\gamma_2}\!(e_2) \,p_{\gamma_1}p_{\gamma_2} \, E ( d_{e_1} d_{e_2} ) \nonumber \\
	&= \sum\limits_{e \in \mE^d_{ij}} A_e^{(i,j)} E ( d_{e}^2 ) + \sum\limits_{(e_1,e_2) \in \mE^u_{ij} \times\mE^u_{ij}} B_{e_1,e_2}^{(i,j)} E ( d_{e_1} d_{e_2} ).   \nonumber
\end{align*}\qed

\medskip
\noindent
{\it Proof of Lemma \ref{lem:A-B}. } 
(i) Let $i,j \in \mV$ and $e \in \mE$. Using Proposition \ref{prop:sumTo1},  $\Gamma_{ij}^{e_1} \subseteq \Gamma_{ij}$ and the definition of the sign function, it follows that
\begin{align}\label{eq:ineqBcoef}
&1 = \sum\nolimits_{\gamma_1 \in \Gamma_{ij}} p_{\gamma_1} \geq \sum\nolimits_{\gamma_1 \in \Gamma^{e_1}_{ij}} p_{\gamma_1} \geq \sum\nolimits_{\gamma_1 \in \Gamma^{e_1}_{ij}}\sgn_{\gamma_1}\!(e_1)  p_{\gamma_1}.
\end{align}
An analogous inequality holds by changing the sign of each term. The same results apply to the pair of nodes $k,\ell\in\mV$. Therefore
\begin{align*}
&-1 \le \big( \sum\nolimits_{\gamma_1 \in \Gamma^{e_1}_{ij}}\sgn_{\gamma_1}\!(e_1)  p_{\gamma_1} \big ) \big( \sum\nolimits_{\gamma_2 \in \Gamma^{e_2}_{k\ell}}\sgn_{\gamma_2}\!(e_2)  p_{\gamma_2} \big ) \le +1,
\end{align*}
hence the statement is proved. (ii) The value of $|B_{e_1,e_2}^{(i,j)(k,\ell)}|$ is $1$ if and only if each of the two factors of the $B$-coefficient assumes the value $\pm 1$. Equivalently, \eqref{eq:ineqBcoef} and the analogous with changed sign, has an equality for both  pairs of nodes $(i,j)$ and $(k,\ell)$. This is verified only under the conditions contemplated in statement (ii), hence (ii) is proved. (iii) Observe that, if two distinct a-paths of $\Gamma_{ij}$ share some edges, they must go through two different outgoing edges of a shared node, thus their path labels differ for at least one edge label in their factorization. An analogous statement holds for a-paths of  $\Gamma_{k\ell}$. Hence there are no terms resulting from the product in \eqref{eq:Bcoef} that can sum to zero. Therefore it holds that $B_{e_1,e_2}^{(i,j)(k,\ell)}= 0$ for all $\alpha_e,\,e \in \mE$, if and only if one of the two factors in \eqref{eq:Bcoef} is equal to zero for all $\alpha_e,\,e \in \mE$ (see \cite[p.~4]{Cox3Ed}). Thus at least one of these factors is equal to zero \cite[p.~510]{Cox3Ed}. This condition is  verified if and only if one or both sets $\Gamma_{ij}^{e_1}$ and $\Gamma_{k\ell}^{e_2}$ do not contain any a-path. \qed

\medskip
\noindent
{\it Proof of Proposition \ref{prop:f2Add}. } 
Assume that all a-paths of $\Gamma_{ij}$ pass through  $k$, and that $i,j,k$ are all distinct, otherwise there is nothing to prove. By Lemma \ref{lem:path-statement3},
$$\Gamma_{ij}=\Gamma_{ik}\times\Gamma_{kj},$$
where $(\gamma,\delta)\in \Gamma_{ik}\times\Gamma_{kj}$ denotes concatenation of $\gamma$ and $\delta$.   By Lemma \ref{lem:ortoDrift}, we have
\begin{align*}
D_{ij} &= \sum\nolimits_{\eta \in \Gamma_{ij}} \Big ( p_\eta \sum\nolimits_{e \in \eta} d^\eta_e\Big ) \nonumber\\
& = \sum\nolimits_{\gamma \in \Gamma_{ik}} \sum\nolimits_{\delta\in\Gamma_{kj}} \Big ( p_\gamma p_\delta \left(\sum\nolimits_{e \in \gamma} d^\gamma_e+\sum\nolimits_{e \in \delta}  d^\delta_e\right)\Big ) \\
&= D_{ik}+D_{kj},
\end{align*}
using Lemma \ref{lem:path-statement2}, the definition of path label and drift. Hence $D_{ij}^2=D_{ik}^2+D_{kj}^2+2D_{ik}D_{kj}$. Note that no edge can be shared between $\gamma\in\Gamma_{ik}$ and $\delta\in\Gamma_{kj}$, as $(\gamma,\delta)\in\Gamma_{ij}$ is an a-path. Similarly, there can at most be one undirected edge in $(\gamma,\delta)$, hence by Lemma \ref{lem:ortoDrift}, $E(D_{ik}D_{kj})=0$, and
$$F_2(i,j)=E(D_{ij}^2)=E(D_{ik}^2)+E(D_{kj}^2)=F_2(i,k)+F_2(k,j).$$

For the reverse statement, assume there exists an a-path $\gamma \in \Gamma_{ij}$ that does not pass through $k$, and assume that all a-paths of $\Gamma_{ij}$ start with an edge of type $i \from i'$, for some $i' \in par(i)$, i.e. they are of type (a,b,c,d) (see Table \ref{tab:path}). Then there is at least one pair of a-paths  $\gamma_1 \in \Gamma_{ik},\,\gamma_2 \in \Gamma_{kj}$ sharing edges. In fact, denote by $\delta_1,\delta_2$ the two subpaths of $\gamma$ from $i$ to $k'$ and from $k'$ to $j$, respectively. Here, $k'$ is the first node not having parents in $\gamma$, so that $\delta_1$ is fixed as an a-path of type (a), whereas $\delta_2$ can be of type (d) or (e). From Proposition \ref{prop:sumTo1}, there is an a-path $\delta_1' \in \Gamma_{k'k}$ that can be of any type. Let $\delta_2' \in \Gamma_{kk'}$ be the reverse a-path of $\delta_1'$. It is possible to verify by inspection that the concatenations $\gamma_1=(\delta_1,\delta_1'),\gamma_2=(\delta_2',\delta_2)$, are two a-paths of $\Gamma_{ik},\,\Gamma_{kj}$, respectively, and share the edges of $\delta_1'$ and $\delta_2'$. With an analogous construction, the same holds for the other possible types of a-paths of $\Gamma_{ij}$.

Since $F_2(i,j)=F_2(i,k)+F_2(k,j)$, then $E(D_{ik}D_{kj})=0$. Note that
\begin{align*}
E(D_{ik}D_{kj}) 	&= \sum\limits_{\substack{\gamma_1 \in \Gamma_{ik} \\ \gamma_2 \in \Gamma_{kj}}} p_{\gamma_1}p_{\gamma_2} \sum\limits_{e \in \mE_{ik} \cap \mE_{kj}} \text{sign}_{\gamma_1}(e)\text{sign}_{\gamma_2}(e) E(d_e^2) \nonumber\\
				&= \sum\limits_{e \in \mE}\Bigg ( \sum\limits_{\gamma_1 \in \Gamma_{ik}} \text{sign}_{\gamma_1}(e)p_{\gamma_1} \Bigg ) \Bigg (  \sum\limits_{\gamma_2 \in \Gamma_{kj}} \text{sign}_{\gamma_2}(e)p_{\gamma_2} \Bigg ) E(d_e^2) 
\end{align*}
where we have used the definition of sign and the linearity of expectation. The expression is linear in $E(d_e^2)$, $e \in \mE$, hence if $E(D_{ik}D_{kj})=0$ for arbitrary (positive) values of $E(d_e^2)$, then the coefficient of $E(d_e^2)$, $e \in \mE$, is zero. The argument in the  proof  of Lemma \ref{lem:A-B}(iii) also applies here to conclude that either of the two factors in the coefficient is zero. Since $\Gamma_{ik}\not=\emptyset$ and  $\Gamma_{kj}\not =\emptyset$ (assuming $i,j,k$ are different), then $\sgn_{\gamma_1}(e)=\sgn_{\gamma_2}(e)=0$, contradicting the existence of a pair of overlapping a-paths.
 Therefore all a-paths of $\Gamma_{ij}$ go through $k$.
 \qed 

\medskip
\noindent
{\it Proof of Proposition \ref{prop:F3configs}. } 
Assume $F_{3-}(i;j,k) \neq 0$. Then there is at least one edge $e \in \mE$ for which $A_{e-}^{(i,j)(i,k)}\neq 0$. It follows that there is a pair of a-paths $(\gamma_1,\gamma_2) \in \Gamma_{ij}^e \times \Gamma_{ik}^e$ such that $\text{sign}_{\gamma_1}(e) \neq \text{sign}_{\gamma_2}(e)$. Let us consider  the possible combinations of two a-paths $(\gamma_1,\gamma_2) \in \Gamma_{ij} \times \Gamma_{ik}$ from Table \ref{tab:path}, and verify for which pairs it is possible to have $\text{sign}_{\gamma_1}(e) \neq \text{sign}_{\gamma_2}(e)$. By inspection, it is immediate to verify that $e$ has the same sign in the combinations (a,a), (a,b), (a,c), (b,d), (d,d) (e,e). Combination (a,d) is not possible, because $i$ would be a root and an admixed node at the same time. Combinations (a,e), (b,e), (c,e) and (d,e) imply that $e \notin \Gamma_{ij}^e \cap \Gamma_{ik}^e$. There remains combinations (b,b), (b,c) and (c,c), that admit one or more shared edges of opposite sign in the two paths. The opposite implication is immediate by definition of $F_{3-}(i;j,k)$ and $A_{e-}^{(i,j)(i,k)}$. \qed 

\medskip
\noindent
{\it Proof of Proposition \ref{prop:heteroWF}.}
We will prove a slightly stronger statement, namely
$$E(V_j(1-V_{j'})) = \sum_{(r,r')\in \mR\times\mR}  D_{r,r'}^{j,j'} E(V_r(1-V_{r'})),  \quad j,j'\in \mV,$$
 by induction on the block index (Appendix A), and where  $D_{r,r'}^{j,j'}$ are non-negative and sum to at most one over $r,r'$ for any $j,j'$. 

First consider nodes $j,j'\in B_1=\mR$. By defining $D^{j,j'}_{r,r'}=1$ if $r=j$, $r'=j'$, and zero otherwise, we have
$$E(V_j(1-V_{j'})) =\sum_{(r,r')\in \mR\times\mR}  D_{r,r'}^{j,j'} E(V_{r}(1-V_{r'})),$$
and the claim holds for  $j,j'\in B_1$. Now assume the statement holds for  nodes in $\cup_{k=1}^m B_k$, $m\ge 1$, and consider $j,j'\in \cup_{k=1}^{m+1} B_k$ such that at least one of $j,j'$ is in $B_{m+1}$, say $j\in B_{m+1}$ (similar if $j'\in B_{m+1}$). If $j\not\in B_1$ then
\begin{align}\label{eq:sumsum}
E(V_{j}(1-V_{j'}))&=E\left(\sum_{i\in\pa(j)} \alpha_{ij}C_{ij}\left(1-\sum_{i'\in\pa(j')} \alpha_{i'j'}C_{i'j'}\right)\right) \\
&= \sum_{(i,i')\in \pa(j)\times\pa(j')} \alpha_{ij}\alpha_{i'j'}E(C_{ij}(1-C_{i'j'})),\nonumber
\end{align}
using that $1=\sum_{i'\in\pa(j')}\alpha_{i'j'}$. If $(i,j)\not=(i',j')$ then $E(C_{ij}(1-C_{i'j'}))=E(V_{i}(1-V_{i'}))$ by conditioning on $(V_{i},V_{i'})$ and using Definition \ref{def:chain}(iii). If $(i,j)=(i',j')$ then $E(C_{ij}(1-C_{ij})=e^{-\tau_{ij}}E(V_{i}(1-V_{i})$, where $\tau_{ij}$ is the population scaled time from $i$ and $j$ \cite{Nei1987}. Define $P(j,j')=\pa(j)\times\pa(j')$. Then
\begin{align*}
E(V_{j}(1-V_{j'})) &= \sum_{(i,i')\in P(j,j')} e^{-\tau_{ij}\delta_{ii'}}\alpha_{ij}\alpha_{i'j'}E(V_{i}(1-V_{i'})) \\
&= \sum_{(i,i')\in P(j,j')} e^{-\tau_{ij}\delta_{ii'}}\alpha_{ij}\alpha_{i'j'}\sum_{(r,r')\in \mR\times\mR} D_{r,r'}^{i,i'} E(V_{r}(1-V_{r'})) \\
&= \sum_{(r,r')\in \mR\times\mR} \left[\sum_{(i,i')\in P(j,j')} e^{-\tau_{ij}\delta_{ii'}}\alpha_{ij}\alpha_{i'j'}  D_{r,r'}^{i,i'} E(V_{r}(1-V_{r'}))\right] \\
&=\sum_{(r,r')\in \mR\times\mR}  D_{r,r'}^{j,j'} E(V_{r}(1-V_{r'}))
\end{align*}
where the second line holds by induction hypothesis  since $i,i'\in\cup_{k=1}^m B_k$, $\delta_{ii'}=1$ if $i=i'$ and zero otherwise, and
$$ D_{r,r'}^{j,j'}=\sum_{(i,i')\in P(j,j')}e^{-\tau_{ij}\delta_{ii'}} \alpha_{ij}\alpha_{i'j'}  D_{r,r'}^{i,i'}. $$
 It is easily seen that the constants above  have the desired properties. If $j\in B_1$ (similar if $j'\in B_1$) then we do the same as above except only $V_{j'}$ is replaced by a sum in \eqref{eq:sumsum}.
 We obtain similarly to \eqref{eq:sumsum},
 \begin{align*}
E(V_{j}(1-V_{j'})) &=\sum_{i'\in \pa(j')}  \alpha_{i'j'}E(V_j(1-C_{i'j'})) \\
&= \sum_{i'\in \pa(j')} \alpha_{i'j'}E(V_j(1-V_{i'})) \\
&= \sum_{i'\in \pa(j')} \alpha_{i'j'}\!\!\!\sum_{(r,r')\in \mR\times\mR} D_{r,r'}^{j,i'} E(V_{r}(1-V_{r'})) \\
&=\sum_{(r,r')\in \mR\times\mR}   D_{r,r'}^{j,j'} E(V_{r}(1-V_{r'})),
\end{align*}
where
$$ D_{r,r'}^{j,j'}=\sum_{i'\in \pa(j')} \alpha_{i'j'}  D_{r,r'}^{j,i'}. $$
If $j'\in B_1$ then the role of $j$ and $j'$ is reversed, yielding a similar answer. In both cases the constants fulfil the desired requirements.
 
Letting $j=j'$ yields the desired result.
\qed

\end{document}